# Hall micromagnetometry of individual two-dimensional ferromagnets

Minsoo Kim[1], Piranavan Kumaravadivel[1,2], John Birkbeck[1,2], Wenjun Kuang[1], Shuigang G. Xu[1,2], D. G. Hopkinson[3], Johannes Knolle[4], P. A. McClarty[5], A. I. Berdyugin[1], Moshe Ben Shalom[1,2], R. V. Gorbachev[1,2], S. J. Haigh[3], Song Liu[6], J. H. Edgar[6], K. S. Novoselov[1,2], I. V. Grigorieva[1], A. K. Geim[1,2]

[1]School of Physics and Astronomy, University of Manchester, Oxford Road, Manchester, M13 9PL, United Kingdom

[2]National Graphene Institute, University of Manchester, Oxford Road, Manchester, M13 9PL, United Kingdom

[3]School of Materials, University of Manchester, Oxford Road, M13 9PL, United Kingdom

[4]Blackett Laboratory, Imperial College London, London SW7 2AZ, United Kingdom

[5]Max Planck Institute for the Physics of Complex Systems, Nöthnitzer Strasse 38, 01187 Dresden, Germany

[6]The Tim Taylor Department of Chemical Engineering, Kansas State University, Manhattan, Kansas 66506, USA

**The recent advent of atomically-thin ferromagnetic crystals has allowed experimental studies of two-dimensional (2D) magnetism[1-9] that not only exhibits novel behavior due to the reduced dimensionality but also often serves as a starting point for understanding of the magnetic properties of bulk materials[10-17]. Here we employ ballistic Hall micromagnetometry[18,19] to study magnetization of individual 2D ferromagnets. Our devices are multilayer van der Waals (vdW) heterostructures[20] comprising of an atomically-thin ferromagnetic crystal placed on top of a Hall bar made from encapsulated[21] graphene. 2D ferromagnets can be replaced repeatedly, making the graphene-based Hall magnetometers reusable and expanding a range of their possible applications. The technique is applied for the quantitative analysis of magnetization and its behavior in atomically thin CrBr₃. The compound is found to remain ferromagnetic down to a monolayer thickness and exhibit high out-of-plane anisotropy. We report how the critical temperature changes with the number of layers and how domain walls propagate through the ultimately thin ferromagnets. The temperature dependence of magnetization varies little with thickness, in agreement with the strongly layered nature of CrBr₃. The observed behavior is markedly different from that given by the simple 2D Ising model normally expected to describe 2D easy-axis ferromagnetism. Due to the increasingly common usage of vdW assembly, the reported approach offers vast possibilities for investigation of 2D magnetism and related phenomena.**

Research on magnetism in strongly layered (vdW) materials is only a couple of years old but has already revealed a number of interesting phenomena including, for example, unexpected changes in magnetic properties as a function of the number of layers[2,17] and the possibility to control magnetism by electric and chemical doping[12-16,22]. Of particular interest are ferromagnetic semiconductors such as $Cr_2Ge_2Te_6$ and $CrI_3$, in which a magnetization-dependent optical response and switching of a magnetization direction by applied electric field have been reported[12-16]. A number of different techniques have been employed to study magnetic properties of the above compounds at a few-layer thickness, including magneto-optical Kerr effect[1,2,15], circular dichroism microscopy[12,14], tunnel magnetoresistance[3-6,16], anomalous Hall effect[7], photoluminescence[8,11] and Raman spectroscopy[9]. Unfortunately, none of those techniques can probe the magnetic field response directly whereas such information is desirable for accurate interpretation of phase transitions, spin arrangements, domain structures and propagation of magnetic walls among others. On the other hand, the conventional magnetometry techniques developed for bulk materials cannot be used to study 2D crystals because of the minute volumes involved: 2D crystals are only a few atoms thick and, typically, several micrometers in size, which makes prospects of their direct magnetization measurements next to impossible[17].

Here we describe the use of ballistic Hall micromagnetometry[18] for magnetization studies of individual 2D ferromagnets. Hall magnetometry technique was initially developed for studying mesoscopic superconductors and ferromagnets, and is based on small Hall probes placed in a vicinity of an object exhibiting a magnetic response[18,19]. In particular, it has been shown that, if electron transport within the probe is ballistic, the measured Hall resistance $R_H$ is given by the total magnetic flux through the central (square) area of the Hall cross. This feature allows not just qualitative observations but quantitative analysis of measured magnetization signals. Whereas the earlier Hall magnetometers were mostly made using a high-mobility 2D electron gas in GaAlAs heterostructures, the recent progress in making vdW heterostructures allows a conceptually straightforward extension of this technology into studies of 2D magnetism. In brief, a ferromagnetic monolayer can be placed on top of a graphene Hall bar using vdW assembly. The use of graphene ensures high sensitivity and low noise measurements, thanks to nm-scale proximity of the ferromagnet to the conducting channel and ballistic transport in graphene up to room temperature[23].

Our Hall bar devices (Fig. 1a) were prepared by encapsulating graphene within hexagonal boron nitride (hBN) using the standard dry transfer method (for details, see Methods). The mesa and electrical contacts were fabricated using electron-beam lithography, dry etching and thin-film metal deposition. A ferromagnetic crystal of interest was mechanically exfoliated in a glovebox filled with pure argon and, to avoid chemical degradation in air[21], sandwiched between a pair of thin hBN crystals inside the oxygen- and water- free atmosphere. The trilayer assembly was then transferred on top of the graphene device so that the 2D ferromagnet covered at least one of the Hall crosses as shown in Fig. 1a (below we refer to such crosses as 'ferromagnet' ones). Each device had some Hall crosses left uncovered to be used as references. Furthermore, we took special precautions to ensure that our graphene devices allowed replacement of the studied 2D samples. To this end, ferromagnetic crystals were transferred in such a way that some part of the encapsulating hBN draped over metal contacts (Fig. 1a and b). Weak adhesion between hBN and gold films allowed us to peel off the trilayer assembly when required. With these arrangements in place, we found that the magnetometers could sustain many replacements of ferromagnetic crystals without degradation in Hall sensitivity (Fig. 1b and Supplementary Fig. S1). Below we focus on magnetic properties of 2D $CrBr_3$, which was chosen primarily because of its high environmental stability as compared to other magnetic 2D materials (e.g., $CrI_3$; see Supplementary Fig. S2)[24]. Indeed, our encapsulated $CrBr_3$ samples remained stable for at least several months without any sign of degradation (Supplementary Fig. S3). Furthermore, to emphasize the generality of the reported approach, Supplementary Information describes experiments, in which the Hall micromagnetometry was applied to another magnetic material $Cr_2Ge_2Te_6$ (Supplementary Fig. S4).

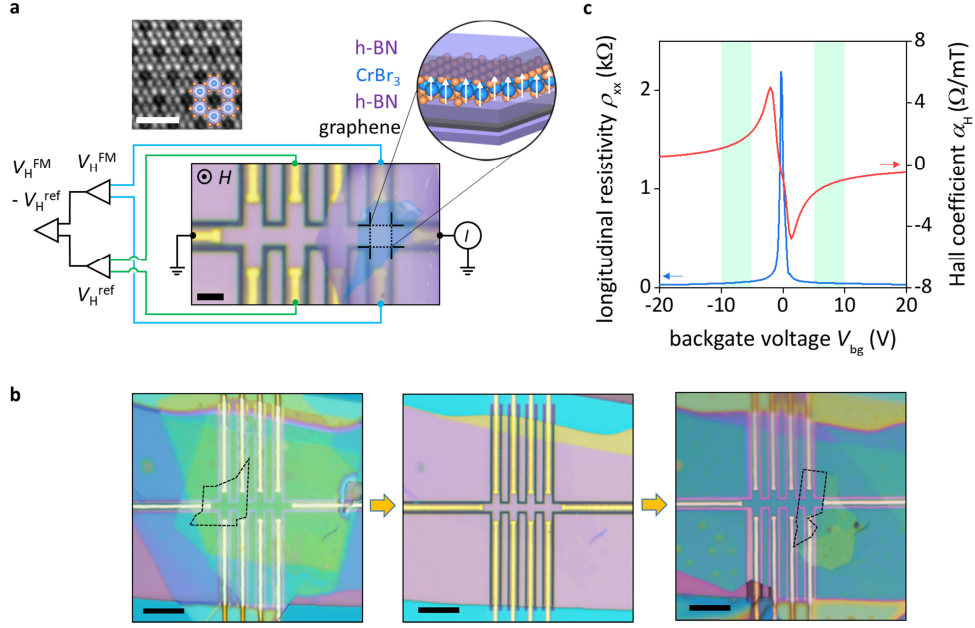

**Figure 1 | Graphene Hall micromagnetometry. a,** Optical micrograph of a graphene Hall bar with encapsulated CrBr$_3$ overlaid on top of one of the crosses. The CrBr$_3$ crystal is shown in false transparent-blue, for clarity. Scale bar, 2 μm. Our measurement circuit using three differential amplifiers is also shown schematically. The FSA is indicated by black lines. Left inset: Transmission electron microscopy image of relatively thick (~10 layers) CrBr$_3$ with its molecular model in the corner. Blue and orange balls denote Cr and Br atoms, respectively. Scale bar, 1 nm. Right inset: Schematic cross-section of the vdW heterostructure. White arrows: spins in monolayer CrBr$_3$. **b,** Reusability of the Hall magnetometry based on van der Waals assembly. An encapsulated monolayer of CrBr$_3$ (left panel) is first removed (center) and then replaced with another encapsulated CrBr$_3$ sample (right). The positions of CrBr$_3$ monolayers are indicated by black lines. Changes in color come from not only the 2D ferromagnets but also a finite thickness of encapsulating hBN crystals that cover large areas. Scale bar, 10 μm. **c,** Examples of $\rho_{xx}$ and $\alpha_H$ as a function of $V_{bg}$ at 2 K. Green areas: Typical range of $V_{bg}$ used for the magnetometry measurements.

An example of basic characterization of our graphene Hall bar devices is provided in Fig. 1c that shows their longitudinal resistivity $\rho_{xx}$ and Hall coefficient $\alpha_H$ as a function of gate voltage $V_{bg}$. The devices were ~2 μm in width and exhibited typical carrier mobilities of ~200,000 cm$^2$ V$^{-1}$ s$^{-1}$, which ensured ballistic transport over a wide temperature (*T*) range. Our experiments followed the same methodology as described previously[18,19]. First, we limited the measurements to magnetic fields below ~0.1 T where the cyclotron radius was larger than the size of our Hall crosses so that the Hall response was linear in field. Next, we analyzed the sensitivity of Hall crosses at different $V_{bg}$ and found that it was maximal for concentrations of ~10$^{12}$ cm$^{-2}$ as indicated in Fig. 1c. The Hall responses of the ferromagnet and reference crosses are given[18,19] by $R_H^{FM} = \alpha_H \cdot (\mu_0 H + B)$ and $R_H^{ref} = \alpha_H \cdot \mu_0 H$, respectively, where $\mu_0 H$ is the externally applied field and $B$ is the average magnetic field within the flux sensitive area (FSA) of a ballistic Hall cross, which is indicated by the dotted black lines in Fig. 2. The differential amplifier scheme shown in Fig. 1a allowed direct measurements of $\Delta = R_H^{FM} - R_H^{ref}$. Because the $\alpha_H$ (that depends on $V_{bg}$) is known experimentally, the found $\Delta = \alpha_H B$ yields directly $B$. Nonetheless, different Hall crosses are always slightly different because of minute changes in geometry and mesoscopic (interference) effects. This leads to non-zero $\Delta$ even for nominally similar

Hall crosses and in the absence of a ferromagnet on top. To suppress this spurious but smooth background (see Fig. 2a), we used high driving currents of the order of 100 μA, which heated graphene's electronic system (but not the nearby 2D ferromagnet) and reduced mesoscopic effects[18,19]. In addition, we measured $\Delta$ at $T$ above the ferromagnetic transition (typically at 40 K for the case of CrBr$_3$; see Methods) and subtracted it from low-$T$ $\Delta$, which allowed us to obtain clean $B$ signals such as those shown in Figs. 2a-d. The achieved resolution in terms of $B$ was about 1 μT, which translates into ~$10^{-3}$ of the magnetic flux quantum. Finally, it is important to note that, because of demagnetization effects, the stray fields from 2D ferromagnets exhibit sharp spikes near their edges, which can reach a value of several mT (Supplementary Fig. S5). This strength is far too small to result in cyclotron orbits within the sub-100 nm extent of the regions and, therefore, the use of the Hall signal as a quantitative measure of the average magnetic flux within the FSA holds[18,19]. Moreover, the spikes are sign changing so that on average they affect little the direction of passing electrons. Accordingly, it is the interior regions of the 2D ferromagnets rather than edges, which contribute mostly to the measured Hall effect, as confirmed in our micromagnetic simulations.

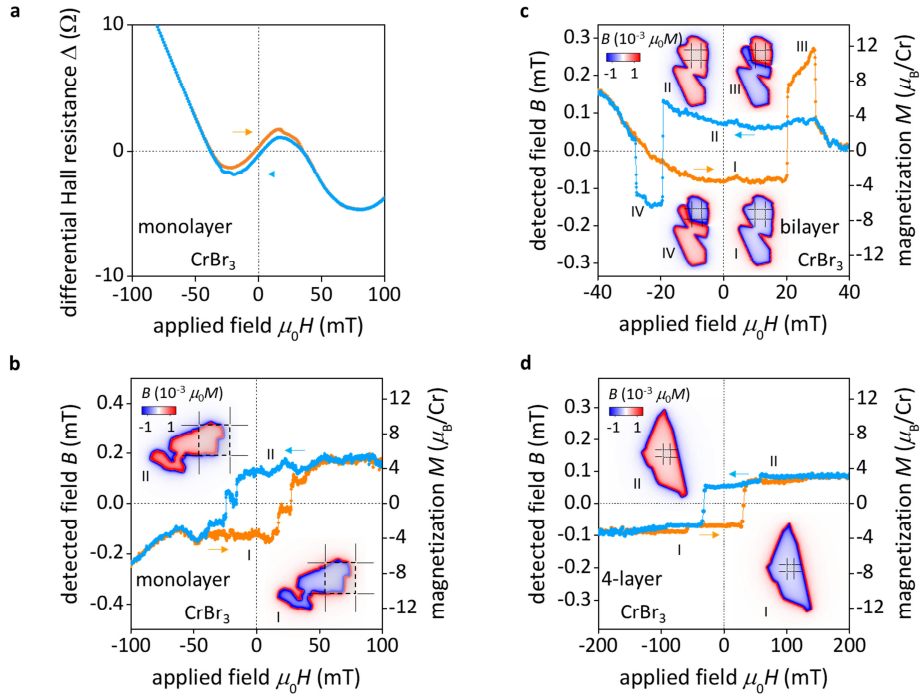

**Figure 2 | Magnetic hysteresis in few-layer CrBr$_3$. a,** Differential Hall resistance $\Delta$ for monolayer CrBr$_3$. It shows magnetic hysteresis at 2 K whereas the smooth background remained practically the same over a wide $T$ range including $T$ above critical. **b-d,** Examples of magnetic hysteresis for mono-, bi- and four- layer CrBr$_3$ crystals at 2 K. (b) shows the same data as in (a) but after subtraction of the $T$-independent background. The insets show the shape of the studied CrBr$_3$ crystals and their positions with respect to the FSA (black lines). The colors represent results of our numerical simulations of the perpendicular field $B(x,y)$ emanating from the 2D ferromagnets into the graphene plane. The average of $B(x,y)$ within the FSA corresponds to the experimentally detected $B$. Color bars: Proportionality coefficient between $B(x,y)$ and $M$ for the given ferromagnets at saturation. The right $y$-axes are to indicate the scale for the saturation magnetization. In (c), possible positions of a domain wall with respect to the FSA are illustrated in the insets marked III and IV.

All the studied CrBr₃ samples exhibited clear hysteresis at low $T$, and the switching field $H$ required to reach the saturation state was typically a few tens of mT (Fig. 2). The value of $B$ at saturation is directly proportional to the magnetization $M$ of a 2D ferromagnet with the proportionality coefficient given by demagnetization effects[25]. This coefficient depends on the shape and size of the measured crystal and can accurately be evaluated using micromagnetic simulations (Methods). Results of such simulations are shown in Figs. 2a-d for the specific crystals under investigation. From the saturation value of $B$ we estimated the magnetization per chromium atom, which is found to be 3.6 ±0.2 $\mu_B$ for all the studied 2D CrBr₃ samples, independently of their thickness (right axes in Figs. 2b-d). This value is in good agreement with 3.8 $\mu_B$ per Cr atom for bulk CrBr₃, which was found using large crystals and a commercial SQUID magnetometer (Supplementary Fig. S6). The agreement corroborates high quality of encapsulated 2D CrBr₃ and the absence of degradation. Importantly, the fact that the remanent magnetization per Cr atom did not depend on the number of layers also proved that spins in different CrBr₃ layers had the same (ferromagnetic) alignment, in contrast to the behavior[2-4,6] observed for 2D CrI₃ where adjacent layers exhibited antiferromagnetism.

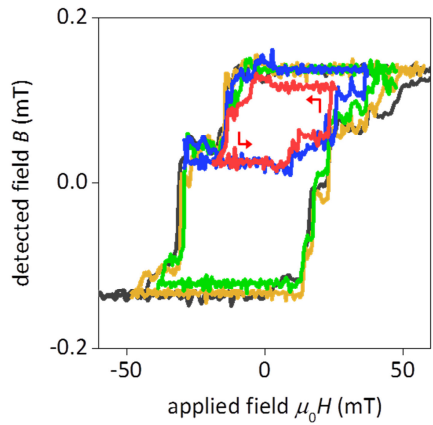

**Figure 3 | Domain wall movements in monolayer CrBr₃.** Hysteresis loops captured by gradually decreasing the sweeping range from 100 mT (black) to 50 (dark yellow), 40 (green), 30 (blue) and 20 (red) mT. Each consecutive loop started in a positive field (for example, the 50 mT loop was recorded after $H$ was swept from -100 mT to +50 mT, then the sweep direction was reversed and the field was swept to -50 mT, and so on). Different constant-$B$ states correspond to different domain configurations inside the monolayer ferromagnet.

The observed hysteresis loops in Fig. 2 exhibit a complex structure, which is particularly profound for the crystal in Fig. 2c where large jumps result in values of $B$ that reach well above the value of the fully spin-polarized state (regions III and IV). The circulation for the resulting (smaller) hysteresis loops is opposite to normal. This counterintuitive behavior is attributed to the formation of magnetic domains within CrBr₃, which can result in a larger flux through the FSA despite the net magnetization of the entire 2D ferromagnetic crystal is smaller. Indeed, our micromagnetic simulations show that, if a domain wall is located near an edge of the FSA (states III and IV in Fig. 2c), the average flux inside the area notably increases with respect to that for states I and II because of weaker demagnetization effects for the two-domain configuration. As the applied field changes further, the domain wall moves away from the FSA region so that crystal's spin polarization gets reversed and $B$ decreases to its normal fully-polarized value determined by the crystal's shape. Similar 'opposite-circulation' loops were observed for other 2D CrBr₃ crystals, although they were usually smaller (e.g., Fig. 2d) as their size was determined by specific domain arrangements with respect to the FSA. In addition, the hysteresis loops always exhibited many small steps such as those seen clearly for the monolayer crystal in Fig. 2b. These are Barkhausen steps that appear because domain walls propagate across the Hall cross through a series of pinned states[26]. For completeness, Fig. 3 shows partial hysteresis loops for the monolayer CrBr₃. In this case, we reversed the sweep direction after a domain wall appeared within the FSA, which allowed us to stabilize the wall in several pinned positions.

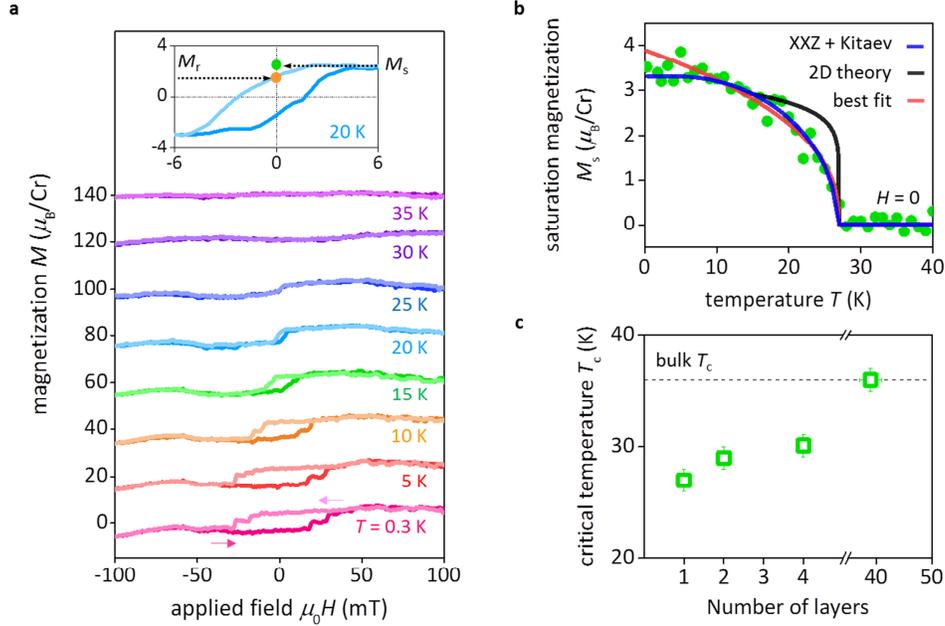

**Figure 4 | Temperature dependence of ferromagnetism in 2D CrBr₃. a,** Magnetic hysteresis in monolayer crystals at several representative $T$. Inset: Magnified coercivity loop at 20 K emphasizes the difference between remanent magnetization $M_r$ and saturation magnetization $M_s$. **b,** The red curve is the best fit to the critical power law, yielding $\beta \approx 0.4$. Blue curve: Mean field description using a spin-anisotropic XXZ model. Black: 2D Ising model critical with $\beta = 0.125$. **c,** Thickness dependence of the critical temperature. $T_c$ for our bulk CrBr₃ is shown by the dashed line.

Next we examine how magnetization of 2D CrBr₃ depends on $T$. Figure 4a shows hysteresis curves found for monolayer CrBr₃. The coercivity rapidly diminishes with increasing $T$ and vanishes above 20 K. Despite the disappearance of hysteresis, a finite magnetization step near zero $H$, which indicates the spin reversal, was observed at higher $T$, up to ~27 K for the case of monolayer CrBr₃. For quantitative analysis, we determined the saturation magnetization $M_s$ from the experimental curves using the procedure illustrated in the inset of Fig. 4a. Figure 4b plots the $T$ dependence of $M_s$ for monolayer CrBr₃. Such plots allowed accurately estimation of the critical (Curie) temperature $T_c$ and, also, more precise evaluation of low-$T$ $M_s$ as compared to the procedure using single-temperature sweeps in Fig. 2. Figure 4c shows $T_c$ as a function of the number of CrBr₃ layers, which gradually decreases from ~36 K for bulk crystals to ~27 K for the monolayer. The relatively small decrease in $T_c$ with decreasing thickness suggests that interlayer magnetic interactions in CrBr₃ are weak so that, in the first approximation, the bulk compound can be considered as a stack of ferromagnetic monolayers.

Empirically, the observed $M_s(T)$ follow the power law $(1\text{-}T/T_c)^\beta$ with the best fit yielding the exponent $\beta \approx 0.4 \pm 0.1$ for monolayer CrBr$_3$ (Fig. 4b; Supplementary Fig. S7). Within our experimental accuracy, all the studied CrBr$_3$ samples exhibited the same $\beta$, independently of the number of layers. This value also agrees with $\beta \approx 0.37$ for bulk CrBr$_3$ as found previously[27]. On one hand, the observation of the $N$-independent critical exponent $\beta$ seems logical because, for a strongly layered material in which ferromagnetism weakly evolves with the number of layers as in CrBr$_3$, $M(T)$ should also change little. On the other hand, strictly-2D ferromagnetism with out-of-plane magnetization is generally described by the Ising model that yields much smaller $\beta = 0.125$ because temperature is less efficient in creating excitations in the 2D space[28]. Such a small exponent matches our data poorly, especially close to $T_c$ where the critical power law scaling behavior should be valid (Fig. 4b; Supplementary Fig. S7). To address the observed disparity, we refer to the work[29] on monolayer CrI$_3$, a sister compound of CrBr$_3$, in which it was suggested that an XXZ model with perturbing bond-dependent Kitaev interactions could be more appropriate to describe magnetic interactions in these layered compounds that are not dissimilar to RuCl$_3$, an archetypal example of a spin liquid[30]. We find that our magnetization data over the entire $T$ range are indeed well described by a mean-field calculation of the above XXZ model (blue curve, Fig 4b). It yields $\beta = 0.5$ at criticality, a value that agrees with the experiment, too (Supplementary Fig. S7). Further work is required to establish the exact nature of magnetic interactions in CrBr$_3$ and whether they indeed lead to mean-field behavior. From this perspective, it is important to mention that, above 20 K, the magnetization loops for 2D CrBr$_3$ are no longer rectangular, and the remanent magnetization $M_r$ is somewhat smaller than $M_s$ (inset of Fig. 4a). This reduction in zero-field magnetization is presumably caused by spontaneous domain formation. To this end, we also measured $M_r(T)$ but found that it was described by the same $\beta$ as $M_s$.

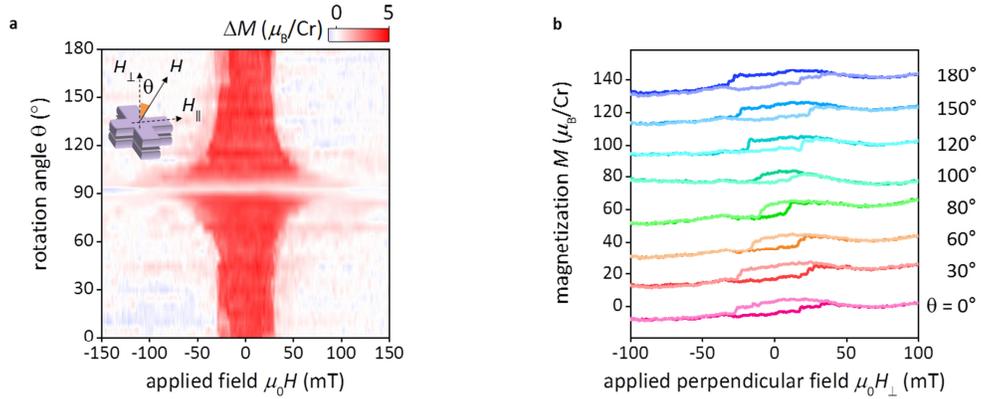

**Figure 5 | Monolayer CrBr$_3$ in tilted magnetic field. a,** Angular dependence of the amplitude of the magnetic hysteresis $\Delta M$ as a function of $H$ and $\theta$. Inset: Our rotation setup is shown schematically. **b,** Hysteresis loops as a function of $H_\perp$ for several angles.

Finally, we studied how hysteresis loops in the 2D ferromagnets change in tilted magnetic fields (Fig. 5a). Interestingly, both $M_s$ and $M_r$ were found to be practically independent of the in-plane field component $H_\parallel$, which corroborates the high uniaxial anisotropy of CrBr$_3$ even in its monolayer form. At the same time, the perpendicular component $H_\perp$ of the external field, which was required to flip the spin polarization, was notably reduced with increasing the tilt angle $\theta$ (Fig. 5b). The likely explanation for the latter behavior is that the in-plane field helps reduce the energy barrier for nucleation of domains of opposite polarity.

To conclude, ballistic Hall micromagnetometry is a powerful tool for investigation of 2D magnetic materials and can reliably measure their magnetization despite minute volumes. The possibility of the controllable sample replacement to test different thicknesses and shapes using the same Hall device greatly enhances the technique's convenience, reliability and experimental possibilities. Our report also shows that Hall micromagnetometry can be used for detecting details of nucleation and propagation of domains walls through 2D materials and for investigation of pinning mechanisms.

## METHODS

**Device fabrication.**     The Hall devices were made from graphene encapsulated between two hBN crystals using the standard dry transfer technique and polypropylene carbonate (PPC) coated polydimethylsiloxane (PDMS) films as stamps. The hBN crystals used in this work were produced by the atmospheric pressure metal-flux method[31]. The metal contacts (5 nm Cr/ 50-70 nm Au) and the Hall bar mesa were fabricated as in the previous reports[32,33]. The only notable difference with respect to the earlier procedures was that, to define the mesa, we etched narrow trenches (100 - 200 nm wide) whereas the rest of the hBN/graphene/hBN stack was kept intact. Also, the thickness of the contact metallization was deliberately chosen close to that of the hBN/graphene/hBN stack. This made the surface of our devices mostly flush to facilitate successive transfers of 2D $CrBr_3$ crystals.

The encapsulated $CrBr_3$ samples were prepared in an argon-filled glovebox with levels of $O_2$ and $H_2O$ below 0.5 ppm. First, bulk $CrBr_3$ (*HQ Graphene*) was exfoliated onto a clean $SiO_2/Si$ substrate. Although we could use bare $SiO_2/Si$ substrates, a thin (7 nm) Au coating was found to increase the probability of finding mono- and bi- layer crystals of $CrBr_3$, in agreement with the earlier work on another layered ferromagnet[7]. Crystals' thickness was first estimated by optical contrast. Then, using a PPC-coated PDMS stamp, a selected $CrBr_3$ crystal was encapsulated between two hBN crystals that were typically 20-40 nm thick. Once encapsulated, the thickness of 2D $CrBr_3$ was verified by atomic force microscopy. Next, again using the argon atmosphere, the encapsulated $CrBr_3$ attached to a PDMS/PPC stamp was aligned and released onto one (or two) of the Hall crosses[34]. The other crosses within the Hall bar devices were left bare for reference measurements. After this, the device was ready for electrical measurements. The etched trenches and weak adhesion between encapsulated $CrBr_3$ and the gold electrodes enabled us to easily peel the $CrBr_3$ stack off the electrode-clamped Hall bar (using PPC/PDMS stamps again) so that $CrBr_3$ of a different thickness can be placed on top of the same Hall magnetometer.

**Transmission electron microscopy.**     Thin $CrBr_3$ was mechanically exfoliated and then encapsulated between two monolayer graphene flakes using the same dry-peel transfer procedures as described above. High angle annular dark field (HAADF) imaging was carried out in a double aberration-corrected JEOL ARM300F with a cold field emission electron gun operating at 80 keV, with a beam convergence semi-angle of 31.74 mrad and HAADF detection range of 68-206 mrad. All aberrations were corrected to better than a $\pi/4$ phase shift at 30 mrad.

**Measurements of magnetic hysteresis.** Hall measurements were performed using the standard lock-in technique at a finite $V_{bg}$ away from the charge neutrality point, where graphene displayed high carrier mobility and, at the same time, strong Hall response. To capture full magnetic hysteresis curves, the Hall resistance was measured at a fixed sensitivity of the lock-in amplifier (Stanford Research Systems 830) while sweeping $H$ to above the switching field. Because the detected $B$ were typically < 0.1% of the switching field, the resolution of digital lock-in amplifiers was insufficient to detect the resulting small changes in Hall voltage. Therefore, we measured the difference in Hall voltages of the 'ferromagnet' and reference Hall crosses using two voltage preamplifiers (Fig. 1a), which allowed the reported high-resolution data.

The excitation current $I_{ac}$ were chosen typically in the range from 10 μA to 200 μA. Higher currents improved the resolution of our measurements by reducing noise and mesoscopic fluctuations in graphene but they could also heat up $CrBr_3$. Therefore, an optimal $I_{ac}$ was selected such that it simultaneously minimized heating effects, mesoscopic fluctuations and noise for a given $V_{bg}$. The devices remained ballistic for all $I_{ac}$. This careful adjustment of experimental parameters allowed a field resolution of about 1 μT. The above sensitivity is comparable to the highest resolution achieved for Hall micromagnetometers based on GaAs/GaAlAs heterostructures[18,19]. We believe that the field resolution can be further improved by using graphene devices with wider contact regions and, therefore, lower contact resistance that is responsible for noise[35].

**Numerical simulations.** Using finite-element analysis, we numerically calculated the spatially varying perpendicular field $B(x,y)$ projected by a ferromagnetic crystal into the graphene plane of magnetometers through hBN spacers. To this end, the exact geometry of 2D $CrBr_3$ crystals (Fig. 2b-d) was obtained by atomic force microscopy. Assuming the constant magnetization $M$ in the out-of-plane direction, we used a numerical mesh in the shape of our $CrBr_3$ crystals and calculated the demagnetizing proportionality coefficient between average $B(x,y)$ within the FSA in the graphene plane and $M$. This coefficient that depended on the shape of the crystal was used to convert the experimentally detected values of $B$ at saturation into $M$.

**Spin-3/2 XXZ-Kitaev model.** We computed the spontaneous magnetization within a local mean field theory for this model described by

$$\mathcal{H}_{XXZ-K} = \sum_{\langle i,j \rangle} \left( \frac{J_\perp}{2} \left( S_i^+ S_j^- + S_i^- S_j^+ \right) + J_\parallel S_i^z S_j^z \right) + K \sum_{\langle i,j \rangle_\alpha} S_i^\alpha S_j^\alpha$$

Decoupling the bilinear exchange and solving the self-consistent equation for the local exchange field at finite $T$, we obtained the average magnetization

$$\langle S_i \rangle = \frac{Tr \left( S_i e^{-\beta \mathcal{H}_{MFT}(i)} \right)}{Z(i)}$$

where $Z(i) = Tr \, e^{-\beta \mathcal{H}_{MFT}(i)}$ with $H_{MFT} = S_i \cdot h_i$. The field $h_i$ is the exchange field from neighboring magnetic sites on the honeycomb lattice. This gives the $T$ dependent magnetization directly. The data do not constrain the parameters of this model except for an overall scale. The corresponding fit in Fig. 4b is for $J_\perp = -5.23$ K, $J_\parallel = -6.33$ K, $K = -1.7$ K and the g-factor $g = 2.2$ defined through $\langle J_{i,Cr} \rangle = g \langle S_i \rangle$.

**References**

1       Gong, C. et al. Discovery of intrinsic ferromagnetism in two-dimensional van der Waals crystals. *Nature* **546**, 265-269 (2017).

2       Huang, B. et al. Layer-dependent ferromagnetism in a van der Waals crystal down to the monolayer limit. *Nature* **546**, 270-273 (2017).

3       Song, T. et al. Giant tunneling magnetoresistance in spin-filter van der Waals heterostructures. *Science* **360**, 1214-1218 (2018).

4       Klein, D. R. et al. Probing magnetism in 2D van der Waals crystalline insulators via electron tunneling. *Science* **360**, 1218-1222 (2018).

5       Ghazaryan, D. et al. Magnon-assisted tunnelling in van der Waals heterostructures based on $CrBr_3$. *Nat. Electron.* **1**, 344-349 (2018).

6       Wang, Z. et al. Very large tunneling magnetoresistance in layered magnetic semiconductor $CrI_3$. *Nat. Commun.* **9**, 2516 (2018).

7       Fei, Z. et al. Two-dimensional itinerant ferromagnetism in atomically thin $Fe_3GeTe_2$. *Nat. Mater.* **17**, 778-782 (2018).

8       Seyler, K. L. et al. Ligand-field helical luminescence in a 2D ferromagnetic insulator. *Nat. Phys.* **14**, 277-281 (2018).

9       Yao, T., Mason, J. G., Huiwen, J., Cava, R. J. & Kenneth, S. B. Magneto-elastic coupling in a potential ferromagnetic 2D atomic crystal. *2D Mater.* **3**, 025035 (2016).

10      McGuire, M. Crystal and magnetic structures in layered, transition metal dihalides and trihalides. *Crystals* **7**, 121 (2017).

11      Zhong, D. et al. Van der Waals engineering of ferromagnetic semiconductor heterostructures for spin and valleytronics. *Sci. Adv.* **3**, e1603113 (2017).

12      Jiang, S., Shan, J. & Mak, K. F. Electric-field switching of two-dimensional van der Waals magnets. *Nat. Mater.* **17**, 406-410 (2018).

13      Huang, B. et al. Electrical control of 2D magnetism in bilayer $CrI_3$. *Nat. Nanotechnol.* **13**, 544-548 (2018).

14      Jiang, S., Li, L., Wang, Z., Mak, K. F. & Shan, J. Controlling magnetism in 2D $CrI_3$ by electrostatic doping. *Nat. Nanotechnol.* **13**, 549-553 (2018).

15      Wang, Z. et al. Electric-field control of magnetism in a few-layered van der Waals ferromagnetic semiconductor. *Nat. Nanotechnol.* **13**, 554-559 (2018).

16      Song, T. et al. Voltage Control of a van der Waals Spin-Filter Magnetic Tunnel Junction. *Nano Lett.* Advance online publication. doi:10.1021/acs.nanolett.8b04160 (2019).

17    Burch, K. S., Mandrus, D. & Park, J.-G. Magnetism in two-dimensional van der Waals materials. *Nature* **563**, 47-52 (2018).

18    Geim, A. K. et al. Phase transitions in individual sub-micrometre superconductors. *Nature* **390**, 259-262 (1997).

19    Novoselov, K. S., Geim, A. K., Dubonos, S. V., Hill, E. W. & Grigorieva, I. V. Subatomic movements of a domain wall in the Peierls potential. *Nature* **426**, 812-816 (2003).

20    Geim, A. K. & Grigorieva, I. V. Van der Waals heterostructures. *Nature* **499**, 419-425 (2013).

21    Cao, Y. et al. Quality heterostructures from two-dimensional crystals unstable in air by their assembly in inert atmosphere. *Nano Lett.* **15**, 4914-4921 (2015).

22    Abramchuk, M. et al. Controlling magnetic and optical properties of the van der Waals crystal $CrCl_{3-x}Br_x$ via mixed halide chemistry. *Adv. Mater.* **30**, 1801325 (2018).

23    Mayorov, A. S. et al. Micrometer-scale ballistic transport in encapsulated graphene at room temperature. *Nano Lett.* **11**, 2396-2399 (2011).

24    Shcherbakov, D. et al. Raman spectroscopy, photocatalytic degradation, and stabilization of atomically thin chromium tri-iodide. *Nano Lett.* **18**, 4214-4219 (2018).

25    Skomski, R., Oepen, H. P. & Kirschner, J. Micromagnetics of ultrathin films with perpendicular magnetic anisotropy. *Phys. Rev. B* **58**, 3223-3227 (1998).

26    Christian, D. A., Novoselov, K. S. & Geim, A. K. Barkhausen statistics from a single domain wall in thin films studied with ballistic Hall magnetometry. *Phys. Rev. B* **74**, 064403 (2006).

27    Ho, J. T. & Litster, J. D. Divergences of the magnetic properties of $CrBr_3$ near the critical point. *J. Appl. Phys.* **40**, 1270-1271 (1969).

28    Vaz, C. A. F., Bland, J. A. C. & Lauhoff, G. Magnetism in ultrathin film structures. *Rep. Prog. Phys.* **71**, 056501 (2008).

29    Xu, C., Feng, J., Xiang, H. & Bellaiche, L. Interplay between Kitaev interaction and single ion anisotropy in ferromagnetic $CrI_3$ and $CrGeTe_3$ monolayers. *NPJ Computat. Mater.* **4**, 57 (2018).

30    Banerjee, A. et al. Neutron scattering in the proximate quantum spin liquid α-$RuCl_3$. *Science* **356**, 1055-1059 (2017).

**Methods-only references**

31    Liu, S. et al. Single crystal growth of millimeter-sized monoisotopic hexagonal boron nitride. *Chem. Mater.* **30**, 6222-622 (2018).

32    Wang, L. et al. One-dimensional electrical contact to a two-dimensional material. *Science* **342**, 614-617 (2013).

33      Ben Shalom, M. et al. Quantum oscillations of the critical current and high-field superconducting proximity in ballistic graphene. *Nat*. *Phys*. **12**, 318-322 (2016).

34      Frisenda, R. et al. Recent progress in the assembly of nanodevices and van der Waals heterostructures by deterministic placement of 2D materials. *Chem*. *Soc*. *Rev*. **47**, 53-68 (2018).

35      Novoselov, K. S. et al. Submicron probes for Hall magnetometry over the extended temperature range from helium to room temperature. *J*. *Appl*. *Phys*. **93**, 10053-10057 (2003).

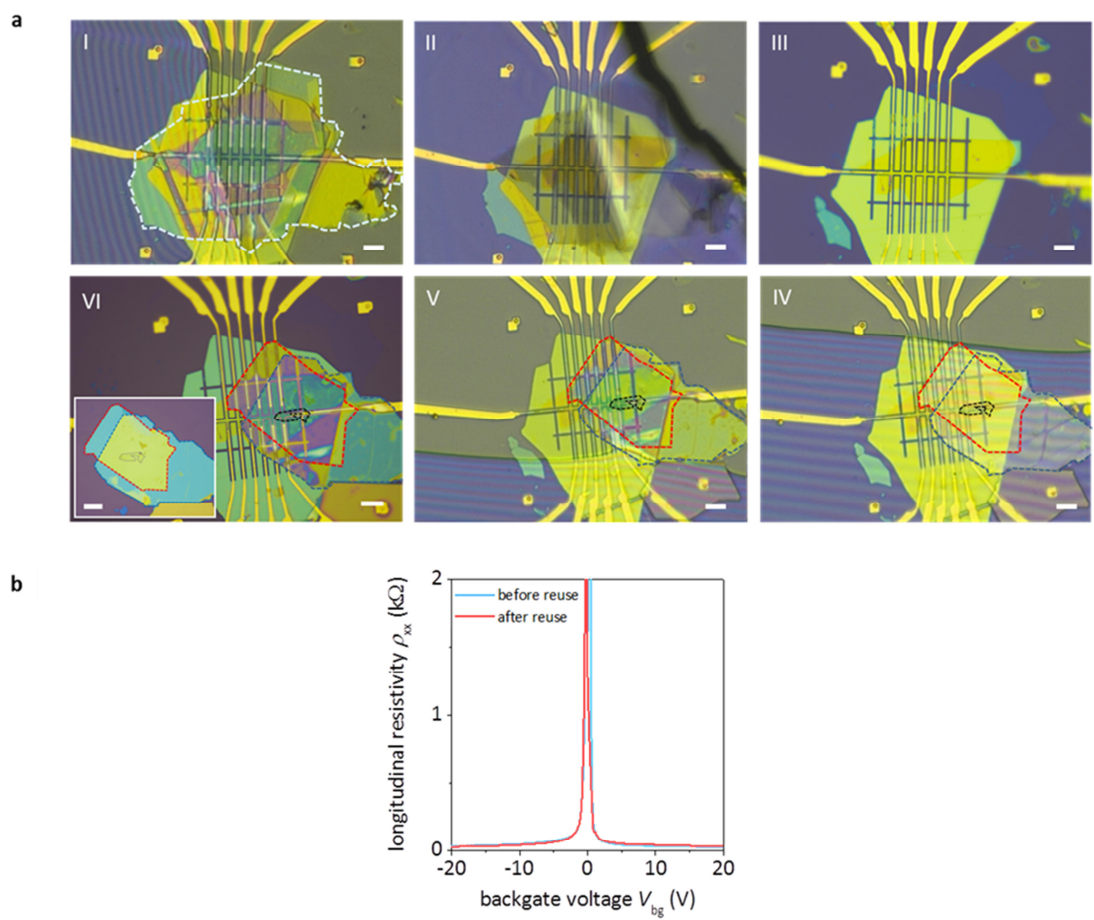

**Figure S1| Reusability of graphene-based Hall micromagnetometers. a,** Optical micrographs during replacement of one 2D ferromagnet with another. Panel **I:** A Hall magnetometer with the already-investigated hBN/2D ferromagnet/hBN sample is brought in contact with a transparent PPC-PDMS stamp at $50^0$C. The dotted lines outline the hBN/2D ferromagnet/hBN stack, parts of which drape over the gold contacts. Panel **II:** The hBN/2D ferromagnet/hBN stack being peeled off the magnetometer and is now seen attached to the PPC-PDMS stamp, still positioned above. Panel **III:** The Hall magnetometer after removal of the old stack is ready for further usage. Panel **IV:** A new encapsulated 2D ferromagnet is aligned above the rightmost Hall cross, after being picked up from a substrate using the PPC-PDMS stamp. Dotted lines: top hBN (red), bottom hBN (blue) and ferromagnetic crystal (black). Panel **V:** The aligned encapsulated 2D ferromagnet is now in contact with the Hall magnetometer heated to $75^0$C but still attached to the PPC/PDMS stamp. Panel **VI:** Finally, the Hall magnetometer with the new encapsulated ferromagnet on top is ready for measurements. The inset shows the same hBN/2D ferromagnet/hBN on the Si/SiO$_2$(285 nm) substrate before its transfer onto the magnetometer. All scale bars, 5 μm. **b,** Longitudinal resistivity $\rho_{xx}$ before (blue) and after (red) the 2D ferromagnet replacement; $T$ = 2 K. The device characteristics change little demonstrating robustness of the graphene-based Hall magnetometers.

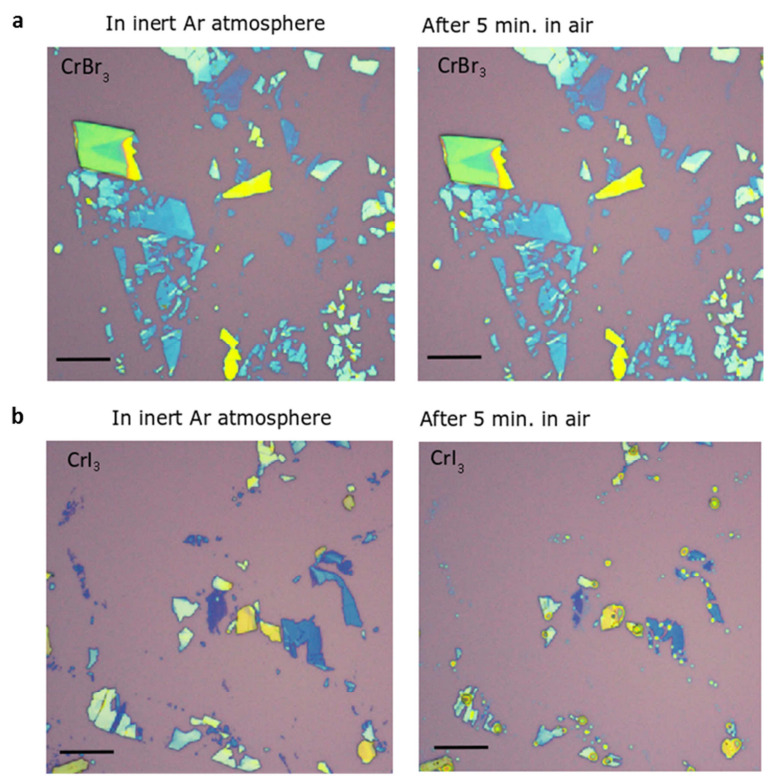

**Figure S2| Stability of magnetic layered trihalides.** Exfoliated $CrBr_3$ (**a**) and $CrI_3$ (**b**) on an oxidized Si substrate (285 nm of $SiO_2$). Left panels: Optical micrographs after exfoliation in an argon-filled glovebox with levels of $O_2$ and $H_2O$ below 0.5 ppm. Right panels: Same are after exposure to ambient air for five minutes. Scale bars, 10 µm. Signs of degradation are clearly visible for $CrI_3$ crystals whereas no changes can be discerned for the case of $CrBr_3$.

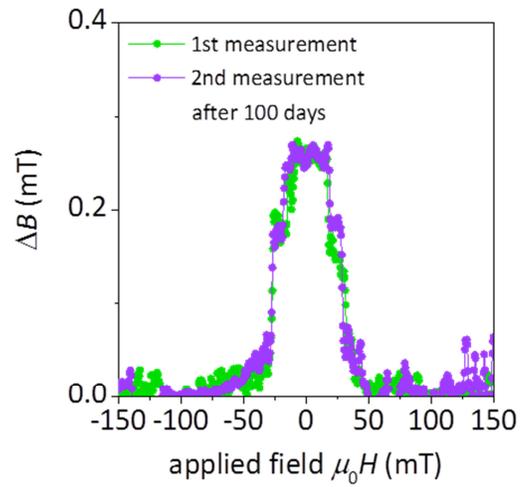

**Figure S3| High stability of encapsulated 2D CrBr₃.** The difference in Hall resistance $R_H$ for up and down sweeps (referred to as the magnetic hysteresis $\Delta B$) for monolayer CrBr₃ at 2 K. After first measurements (green), the encapsulated ferromagnet was stored for ~100 days and then measured again (purple). The reproducibility of $\Delta B$ shows little changes in encapsulated CrBr₃. The small differences in the size of the Barkhausen steps[1] probably comes from slight changes in domain wall positions after the thermal cycling or with time.

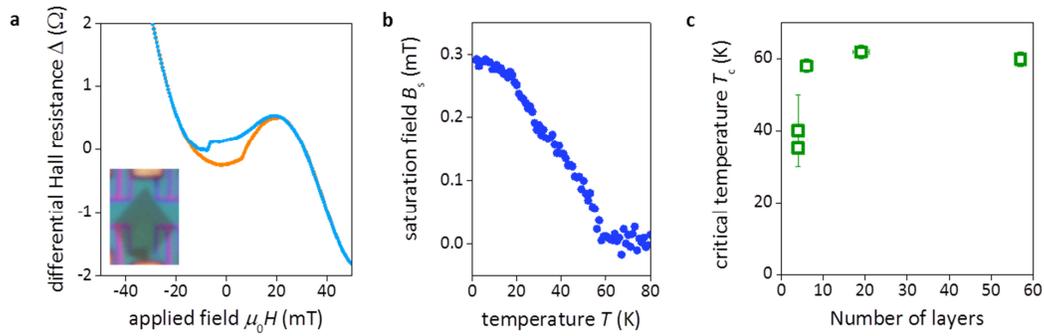

**Figure S4| Ferromagnetism of 2D Cr₂Ge₂Te₆ probed by ballistic Hall micromagnetometry. a**, Differential Hall resistance Δ of six-layer Cr₂Ge₂Te₆ exhibiting magnetic hysteresis; $T$ = 2 K. Inset: Optical micrograph of the encapsulated graphene Hall device with the hBN/encapsulated Cr₂Ge₂Te₆/hBN visible on top. **b**, Temperature dependence of the detected saturation field $B_s$ for this 2D crystal. The curve allows an estimate for $T_c$ as 58±2 K. As described in the previous study[2] on Cr₂Ge₂Te₆, the material has weak out-of-plane anisotropy so that $B_s$ decreases rather slowly with $T$. **c**, Layer dependence of $T_c$ shows a rapid decrease with decreasing thickness, in agreement with ref. 2.

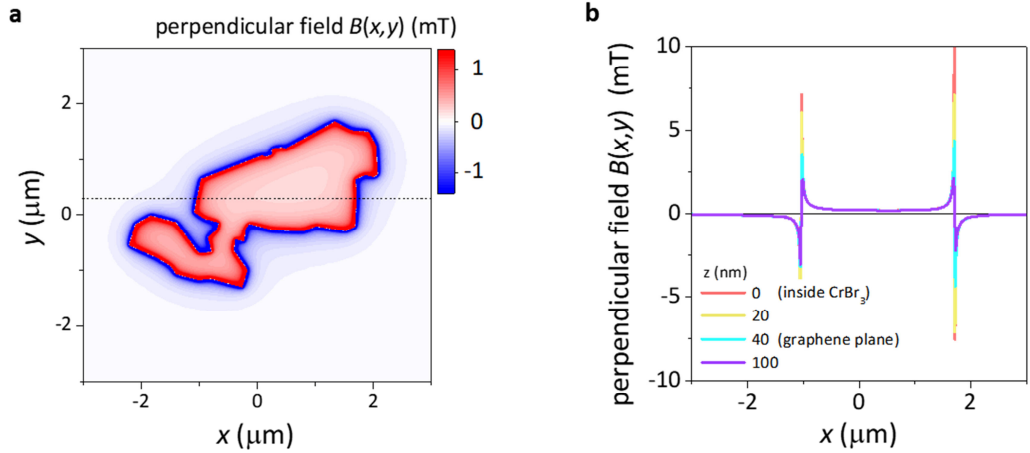

**Figure S5| Simulated magnetic field from a monolayer CrBr₃ crystal. a,** Example of numerical simulations of a perpendicular component of magnetic field $B(x,y)$. Monolayer CrBr₃ (shape of the sample from Fig. 2b of the main text) is fully polarized with 3.8 $\mu_B$ per Cr atom. **b,** $B$ along the dashed line in (a) at different vertical distances from the CrBr₃ plane (color coded). The field profile at 40 nm corresponds to the color map in (a) and was used in the analysis of the sample in Fig. 2b of the main text. Despite the large spikes in $B$ near the edges, the dominant contribution to the Hall signal (total flux within the FSA) comes from the central part of the 2D ferromagnet because of 1) a greater spatial extent of the latter region and 2) the sign-changing field near the edges, which averages to near zero.

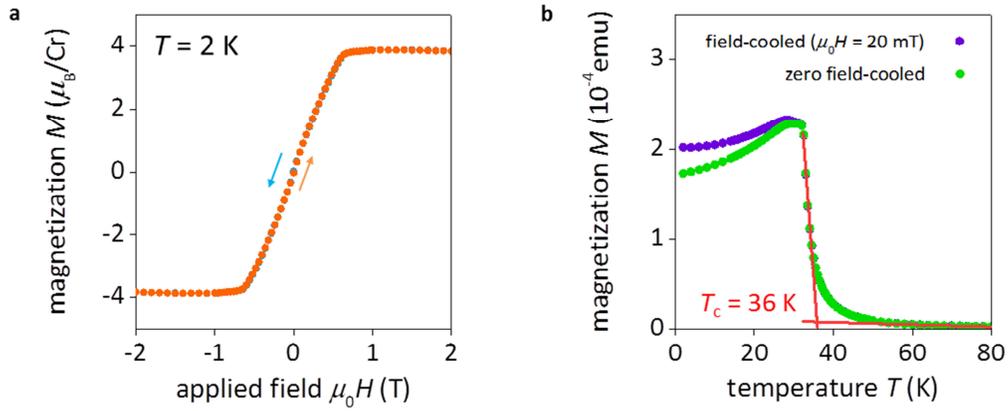

**Figure S6| SQUID measurements of bulk CrBr₃. a,** Magnetic hysteresis curves measured using a CrBr₃ crystal of ~1×1×0.1 mm³ in size. The magnetic field $H$ was applied perpendicular to CrBr₃ layers. The SQUID was from *Quantum Design*. Saturation magnetization corresponds to ~3.8 $\mu_B$ per Cr atom. **b,** Temperature dependence of magnetization of the CrBr₃ crystal using zero field-cooled and 20 mT-field-cooled regimes (green and violet curves, respectively). The Curie temperature $T_c$ is close to 36 K, in agreement with in the previous reports[3].

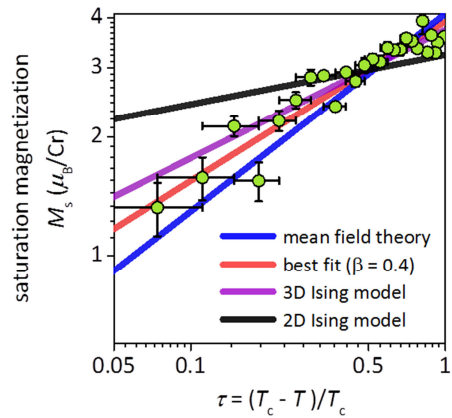

**Figure S7| Power law fits to the magnetization data for monolayer CrBr₃.** The saturation magnetization $M_s$ (data of Fig. 4b in the main text) is plotted against $\tau = (T_c - T)/T_c$ with $T_c = 27$ K. The red line is the best fit yielding $\beta = 0.4$; blue line is a fit using the mean-field exponent $\beta = 1/2$; the purple – $\beta = 0.325$ from the 3D Ising model; and black – the 2D Ising exponent $\beta = 0.125$. Within our experimental accuracy, all the critical power laws can describe the data close to $T_c$, except for the 2D Ising model. The blue and purple fits can be improved by slightly adjusting the assumed $T_c$.

**Supplementary references**

1    Christian, D. A., Novoselov, K. S. & Geim, A. K. Barkhausen statistics from a single domain wall in thin films studied with ballistic Hall magnetometry. *Phys. Rev. B* **74**, 064403 (2006).

2    Gong, C. et al. Discovery of intrinsic ferromagnetism in two-dimensional van der Waals crystals. *Nature* **546**, 265-269 (2017).

3    Tsubokawa, I. On the magnetic properties of a CrBr₃ single crystal. *J. Phys. Soc. Jpn*. **15**, 1664-1668 (1960).